\documentclass[fleqn,usenatbib]{mnras}
\usepackage{newtxtext,newtxmath}
\usepackage[T1]{fontenc}
\usepackage{graphicx}
\usepackage{anyfontsize}
\usepackage{tabularx}
\usepackage{supertabular}
\usepackage[math]{cellspace}

\usepackage{xspace}
\usepackage[dvipsnames]{xcolor}
\usepackage{fontawesome}
\usepackage{multirow}
\usepackage[authoryear]{natbib}
\usepackage{orcidlink}
\usepackage[super]{nth}

\usepackage{hyperref}
\hypersetup{colorlinks=true, allcolors=Blue}

\newcommand{\LUPM}{%
Laboratoire Univers et Particules de Montpellier, CNRS \& Université de Montpellier, Parvis Alexander Grothendieck, Montpellier, France 34090
}

\newcommand{\Newcastle}{%
School of Mathematics, Statistics and Physics, Newcastle University, Herschel Building, Newcastle-upon-Tyne, NE1 7RU, UK
}

\newcommand{\DurhamICC}{%
Department of Physics, Institute for Computational Cosmology, Durham University, South Road, Durham DH1 3LE, UK
}

\newcommand{\DurhamCEA}{%
Department of Physics, Centre for Extragalactic Astronomy, Durham University, South Road, Durham DH1 3LE, UK
}

\newcommand{\Northeastern}{%
Department of Physics, Northeastern University, 360 Huntington Ave, Boston, MA USA
}
\newcommand{\Liege}{%
STAR Institute, Quartier Agora - All\'ee du six Ao\^ut, 19c B-4000 Li\`ege, Belgium}

\newcommand{\Aalto}{%
Department of Computer Science, Aalto University, PO Box 15400, Espoo, FI-00 076, Finland
}
\newcommand{\Helsinki}{%
Department of Physics, Faculty of Science, University of Helsinki, 00014-Helsinki, Finland}

\newcommand{\UTAustin}{%
Department of Astronomy, The University of Texas at Austin, Austin, TX, USA
}

\newcommand{\DAWN}{%
Cosmic Dawn Centre (DAWN), Denmark
}

\newcommand{\NBI}{%
Niels Bohr Institute, University of Copenhagen, Jagtvej 128, 2200 Copenhagen, Denmark}

\newcommand{\JPL}{%
Jet Propulsion Laboratory, California Institute of Technology, 4800, Oak Grove Drive, Pasadena, CA, USA}

\newcommand{\PMO}{Purple Mountain Observatory, Chinese Academy of Sciences, 10
Yuanhua Road, Nanjing 210023, China}

\newcommand{\IAP}{Institut d’Astrophysique de Paris, UMR 7095, CNRS, Sorbonne Universit\'e, 98 bis boulevard Arago, F-75014 Paris, France}

\newcommand{\LAM}{Laboratoire d'astrophysique de Marseille, Aix Marseille University, CNRS, CNES, Marseille, France}

\newcommand{\UCSB}{Department of Phyiscs University of California Santa Barbara, CA,93106, CA}

\newcommand{\Rochester}{Laboratory for Multiwavelength Astrophysics, School of Physics and Astronomy, Rochester Institute of Technology, 84 Lomb Memorial Drive, Rochester, NY 14623, USA}

\newcommand{\STScI}{Space Telescope Science Institute, 3700 San Martin Drive, Baltimore, MD 21218, USA}

\newcommand{\UCSC}{Department of Astronomy and Astrophysics, University of California, Santa Cruz, 1156 High Street, Santa Cruz, CA 95064 USA}

\newcommand{\NAOJ}{National Astronomical Observatory of Japan, 2-21-1, Osawa, Mitaka, Tokyo, Japan}

\newcommand{\Hawaii}{Department of Physics and Astronomy, University of Hawaii, Hilo, 200 W Kawili St, Hilo, HI 96720, USA}

\newcommand{\Caltech}{Caltech/IPAC, 1200 E. California Blvd., Pasadena, CA 91125, USA}

\newcommand{\DTU}{DTU-Space, Technical University of Denmark, Elektrovej 327, DK-2800 Kgs. Lyngby, Denmark}

\title[COWLS II: 17 spectacular strong lenses]{The COSMOS-Web Lens Survey (COWLS) II: depth, resolution, and NIR coverage from \textit{JWST} reveal 17 spectacular lenses}

\author[G.\ Mahler et al.]{Guillaume Mahler\,$\!^{1,2,3}$\thanks{E-mail: guillaume.mahler@uliege.be}\orcidlink{0000-0003-3266-2001},
James W.\ Nightingale\,$\!^{4}$\orcidlink{0000-0002-8987-7401}, 
Natalie B. Hogg$^{5}$\orcidlink{0000-0001-9346-4477}, 
Ghassem Gozaliasl$^{6,7}$,
\newauthor
Jacqueline McCleary$^{8}$,
Qiuhan He$^{2,3}$\orcidlink{0000-0003-3672-9365},
Edward Berman$^{8}$,
Maximilien Franco$^{9}$,
Daizhong Liu$^{10}$,
\newauthor
Richard J.\ Massey$^{2,3}$\orcidlink{0000-0002-6085-3780},
Wilfried Mercier$^{11}$,
Diana Scognamiglio$^{12}$,
Marko Shuntov$^{13,14}$,
\newauthor
Maximilian von Wietersheim-Kramsta$^{2,3}$, 
Louise Paquereau,$^{15}$
Olivier Ilbert,$^{16}$
Natalie Allen,$^{12, 13}$
\newauthor
Sune Toft,$^{12,13}$
Hollis B. Akins,$^{9}$
Caitlin M. Casey,$^{9,17,13}$
Jeyhan S. Kartaltepe,$^{18}$
Anton M. Koekemoer,$^{19}$
\newauthor
Henry Joy McCracken,$^{15}$
Jason D. Rhodes,$^{12}$
Brant E. Robertson,$^{20}$
Jorge A. Zavala,$^{21}$
Nicole E. Drakos,$^{22}$
\newauthor
Andreas L. Faisst,$^{23}$
Georgios E. Magdis,$^{13,24,14}$
and Shuowen Jin$^{13,24}$
\\
Affiliations can be found after the references.
}

\date{}

\pubyear{\the\year{}}

\begin{document}
\label{firstpage}
\pagerange{\pageref{firstpage}--\pageref{lastpage}}
\maketitle

\begin{abstract}
The COSMOS-Web Lens Survey (COWLS) presents the first systematic search for strong gravitational lenses in the COSMOS-Web field using data from the \textit{James Webb} Space Telescope (\textit{JWST}). Using high-resolution NIRCam imaging, we visually inspected over 42\,660 galaxies and identified over 400 lensing candidates. From this sample and based on \textit{JWST}/NIRCam imaging only, we report here the 17 most obvious and spectacular strong lensing systems. These lenses, characterised by large Einstein rings and arcs and their distinct lens and source colours, were found through only the visual inspection of the lens-light-subtracted image data and were immediately visible due to their spectacular appearance. We showcase how spectacular strong lenses are at the extremes of lens parameter space. Their exceptionally high signal-to-noise, multi-wavelength imaging enables unprecedented lensing analysis, including `\textit{HST}-dark' source galaxies that are also invisible in the deeper bluer \textit{JWST} wavebands, enabling clean deblending between the lens and the source. Sources may exhibit dramatic morphological changes across wavelengths, and dust absorption within lenses may be detectable by eye. No other instrument, including the \textit{Hubble} Space Telescope, can discover or image such lenses with comparable detail. We estimate that \textit{JWST} uncovers a new spectacular lens approximately every 10 to 12 NIRCam pointings, suggesting that over 40 such lenses remain undetected within its first three years of observations.
All COWLS data is publicly available on GitHub \href{https://github.com/Jammy2211/COWLS_COSMOS_Web_Lens_Survey}{\faGithub}.
\end{abstract}



\section{Introduction}

\begin{figure*}
 \centering
    \includegraphics[width=\textwidth]{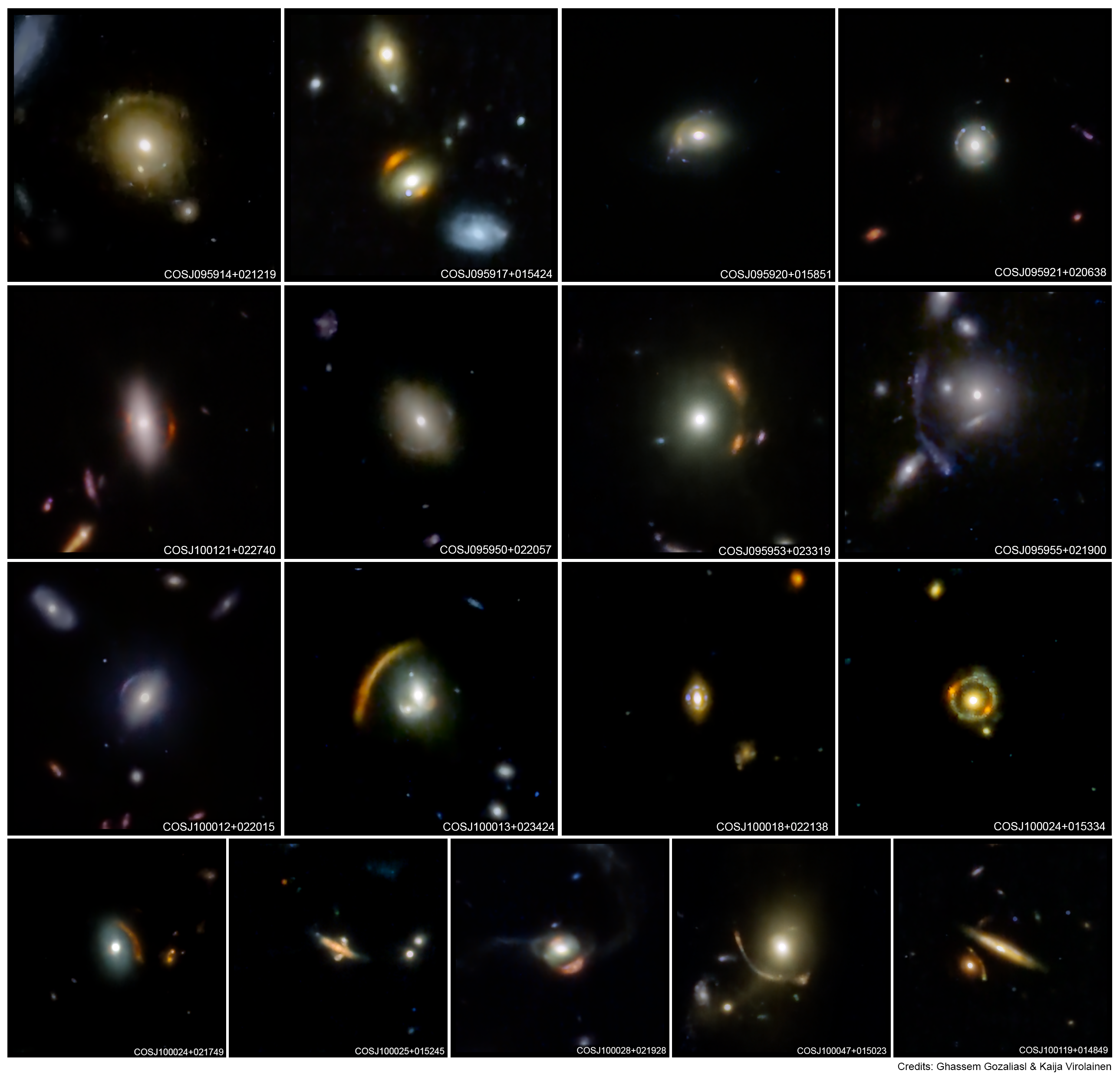}
    \caption{The 17 most spectacular lenses from the COWLS sample, revealed by the \textit{JWST} imaging through our visual inspection of the COSMOS-Web field. Each panel is oriented North up and East left with a negative 20 degree angles (coming from the native orientation of the image). The size of the field of view is adapted for the rendering and can be estimated from \autoref{fig:JWSTvsHST}. The images are produced combining the four filters (F115W, F150W, F277W, F444W) for an ideal rendering of the lensing evidence.}
    \label{fig:lens}
\end{figure*}

Observations of strong gravitational lensing enable a wealth of science. The underlying mass distribution of the deflectors can be probed \citep{Auger2010,Oguri2014,Sonnenfeld2015,VandeVyvere2022, Etherington2023}, the stellar-to-halo mass relation and the initial mass function of galaxies can be constrained \citep{Wang2024, Smith2015}, and substructure present in the halo of the deflector or along the line of sight can be detected \citep{Vegetti2012,Hezaveh2016,Nierenberg2020, Nightingale2024}. Strong lensing systems enable detailed study of distant, magnified sources, from star-forming galaxies \citep{Jones2013,Zavala2018} to Type Ia supernovae \citep{Goobar:2016uuf, Goobar:2022wan} and quasars \citep{Sluse2019}; the latter two are particularly significant as they provide a cosmological model independent measurement of the Hubble parameter $H_0$ via the delay between the arrival times of the multiple images \citep{Refsdal1964, Treu2016,Birrer2020,Birrer2024}. 

All of this science relies on the search for lenses. These searches can commence with ground-based observations, or directly from space. Often, promising lens candidates identified in ground-based surveys are followed up with imaging from space for higher spatial resolution; the most prominent example of this strategy is the Sloan Lens ACS (SLACS) survey \citep{Bolton2006}, which followed-up candidates identified in the Sloan Digital Sky Survey (SDSS) with the \textit{Hubble} Space Telescope (\textit{HST}). Commencing directly from space, the COSMOS survey imaged a 1.64 deg$^2$ area of sky with \textit{HST}, making it, at the time, the largest contiguous high-resolution astronomical imaging survey ever performed from space \citep{Scoville:2006vq,Koekemoer2007}. The COSMOS \textit{HST} data revealed 271 strong gravitational lens candidates \citep{Faure2008, Jackson2008, Pourrahmani2018}, however it is ambiguous which are truly lenses, indicating a high false positive rate.

More recently, the \textit{Euclid} space mission \citep{Scaramella2022,Mellier2024} starting surveying 15\,000 deg$^2$ with a resolution that closely matches that of \textit{HST}. Such a large sky coverage enables the discovery of very nearby lenses \citep{ORiordan2025}. The \textit{Euclid} early release observations \citep{Cuillandre2024}, already yielded numerous candidate based on visual inspection \citep{Acevedo2024} and machine learning detection algorithms \citep{Pearce-Casey2024,Nagam2025}. Over the entire 15\,000 deg$^2$ of the \textit{Euclid} survey, about $\mathcal{O}(10^5)$ strong lenses are expected to be observed \citep{Collett:2015roa, Holloway:2023axl, Ferrami:2024obm}.

As part of the COSMOS-Web survey \citep{Casey2023}, \textit{JWST} has observed a 0.54 deg$^2$ subset of the COSMOS field using the NIRCam and MIRI instruments\footnote{The MIRI observations cover a 0.19 deg$^2$ area.}. We present results from the COSMOS-Web Lens Survey (COWLS), where in this letter, COWLS II, we demonstrate how `spectacular' strong lenses are immediately recognisable by eye in \textit{JWST} NIRCam imaging with minimal preprocessing. Starting with deep \textit{JWST} imaging uncovers lenses at the extremes of lensing parameter space, that ground-based or wide-field imaging based selection (e.g. \textit{Euclid}) will miss. NIRCam provides unmatched resolution and signal-to-noise over four wavebands, which we show enables unique studies of high redshift source galaxies, detailed lens mass modelling and even dust absorption. Comparison with COSMOS F814W \textit{HST} imaging shows that only a fraction of these lenses would be confidently identified as strong lenses from \textit{HST}, and none would offer the multi-wavelength data essential for rich scientific study. We found our 17 spectacular strong lenses in just 0.54 deg$^2$ of COSMOS-Web NIRCam data, we therefore estimate how many unreported spectacular lenses \textit{JWST} has observed over three years since launch.

In COWLS Paper I \citep{Nightingale2025}, we present the inspection campaign using the automated lens modelling software \texttt{PyAutoLens}~\href{https://github.com/Jammy2211/PyAutoLens}{\faGithub} \citep{Nightingale2015, Nightingale2018, pyautolens}. This campaign identifies over 100 highly ranked candidates, including the highest redshift lenses ($z > 2$) and sources ($z > 5$) known to date. All lenses are located within the $0.54$\,deg$^2$ COSMOS-Web field, which allows for joint strong and weak lensing analyses \citep{Birrer2017, Fleury2021, Hogg2022, Hogg:2025wac}. A subset of these lenses have small lens-source separations, providing the potential for measuring supermassive black hole masses \citep{Nightingale2024}. In COWLS Paper III \citep{Hogg2025b}, we compare the observed properties and abundance of COSMOS-Web strong lenses with predictions based on the COWLS catalogue. One spectacular lens, the COSMOS-Web Ring \citep{Mercier2024, VanDokkum2024}, was recently spectroscopically confirmed to be the highest redshift known galaxy-scale lens galaxy ($z = 2.02$) and have a $z = 5.10$ source galaxy \citep{Shuntov2025}, highlighting the unique nature of the lenses presented in this letter.

This work is organised as follows: in \autoref{sec:data} we describe how our data was taken and catalogued; in \autoref{sec:results} we present the spectacular strong lenses and the method used to identify them; in \autoref{sec:discussion} we discuss our results and conclusions. All magnitudes are given in the AB system \citep{Oke1974}.

\section{The COSMOS-Web data} \label{sec:data}

\subsection{Imaging}
The \textit{JWST} Cycle 1 program COSMOS-Web (GO\#1727, PI: Casey \& Kartaltepe, \citealt{Casey2023}), is a photometric survey that consists of imaging in four NIRCam  filters (F115W, F150W, F277W, F444W) and one MIRI filter (F770W). The NIRCam (MIRI) filters reach a $5\sigma$ point source depth of AB mag 27.2$-$28.2 (25.7), measured in empty apertures of 0.15 arcsec (0.3 arcsec) radius. 

\subsection{Photometric catalogue}
The COSMOS-Web team produced a unified photometric catalogue including an automated morphological analysis, based on \texttt{SourceExtractor++} photometric software \citep{Kummel2020,Bertin2020}, which is fully described in \cite{Shuntov2024}. We provide here the relevant characteristics of the catalogue used for our selection. The initial morphological analysis includes a morphological profile fit of the galaxy in each band with one or two S\'ersic profiles. In addition, the software produces a stellarity coefficient ranging from 0 to 1 based on how close to a point-like feature the detection is; for example, an object with a stellarity of 1 is star-like. This process yields a robust photometry for galaxies, and an image with the model of the light distribution derived from the fitted profile. This model is then subtracted from the original image and visually inspect for deviations from the profile. All three types of image (original, model image and model-subtracted images) are used during our search for strong lenses. 

\subsection{Selection of objects to inspect}
Starting from this photometric catalogue presented in \cite{Shuntov2024}, we selected all galaxies brighter than 23 AB mag in F277W and with a stellarity strictly below 1. A galaxy with an apparent magnitude of 23 is relatively faint; we chose this as a conservative cut-off to select for objects massive enough to be strong lenses whilst still imposing a reasonable limit on the total number of objects to visually inspect. This selection resulted in 42\,660 objects to be inspected for signs of strong lensing. A magnitude cut of 23 in the F277W band represents a median and \nth{68} percentile stellar mass of $(3.8\pm 3.3)\times 10^8 M_{\odot}$, $(2.9\pm 2.6)\times10^9M_{\odot}$, and  $(6\pm6)\times10^9M_{\odot}$ at $z=0.3\pm0.05$, $z=1.0\pm0.1$, and $z=2.0\pm0.1$. For the estimated number of galaxies acting as a lens within this parameter space, we refer the reader to COWLS paper III \cite{Hogg2025b}.

\section{Spectacular strong lenses in COSMOS-Web} \label{sec:results}

The visual inspection of the COSMOS-Web field was carried out in two stages and is fully described in the overview paper (COWLS I; \citealt{Nightingale2025}). We here provide a short description of the process which led to the systems highlighted in this paper. In the first stage, inspectors examined the original image, model image and model-subtracted image of each object. This round of inspection relied on a rapid visualisation of all 42\,660 objects, designed to yield a decision by each inspector within a few seconds. Each object was presented to the inspectors with six different images: the image data, model and model-subtracted images in two different filter combinations: F115W/F150W/F444W and F115W/F277W/F444W. Using the model-subtracted image meant inspectors were better able to spot lensed light closer to the central part of the lens, usually much brighter in the detection image.

Each galaxy was classified by at least four inspectors as: ``Yes, this is a lens'', ``No, this is not a lens'', ``Maybe this is a lens'', ``Revisit later'', and ``User category''. Revisit later offered the inspectors the chance to go back and spend more time on the inspection of ambiguous objects, and the ``User category'' allowed inspectors to flag galaxies of interest beyond the search for strong lenses. The data for the first round of visual inspection was split in half, with the first half inspected by four people (GM, JWN, QH, JMcC) and the second half by five people (GM, JWN, NBH, QH, JMcC). 

From the initial 42\,660 objects, 32 objects were flagged as ``Yes'' by more than $50\%$ of inspectors. These candidates were reviewed again by six members of the COSMOS-Web Lensing Working Group (GM, JWN, NBH, QH, MvWK, RJM), who each flagged which systems they thought were the most obvious lenses, showing dramatic lensing evidence. We present in \autoref{fig:lens} the 17 most spectacular systems where a majority of inspectors confidently flagged them as such. This selection was solely based on NIRCam imaging, and we discuss in the next section the specific features which support this ranking. 

\begin{table*}
\centering
\begin{tabular}{Scccccccl}
COWLS ID & ID (\autoref{fig:JWSTvsHST}) & RA [deg] & Dec [deg] & $\theta_{\rm E}$ [arcsec] & $z_{\rm lens}$ &$z_{\rm source}$ & Identification\\ 
\hline
\hline
COSJ095914+021219 &A& 149.811421524 & 2.205440107 & 1.48 & 1.053$^{\rm a}$ & & \protect\cite{Faure2008} \\ 
COSJ095917+015424 &B& 149.821462195 & 1.906873033 & 0.58 & 1.252$^{\rm b}$ & &This work\\ 
COSJ095920+015851 &C& 149.833887486 & 1.980927007 & 0.45 & 0.974$^{\rm a}$ & & This work\\ 
COSJ095921+020638 &D& 149.840689409 & 2.110652506 & 0.71 & 0.469$^{\rm a}$ & &\protect\cite{Faure2008}\\ 
COSJ095950+022057 &E& 149.961430320 & 2.349411544 & 1.04 & 0.939$^{\rm a}$ & &\protect\cite{Pourrahmani2018} \\ 
COSJ095953+023319 &F& 149.974678146 & 2.555445653 & 1.44 & 0.731$^{\rm a}$ & &This work\\ 
COSJ095955+021900 &G& 149.983162054 & 2.316880570 & 2.11 & 0.577$^{\rm b}$ & & \protect\cite{More2012}\\ 
COSJ100012+022015 &H& 150.052588227 & 2.337706469 & 0.75 & 0.377$^{\rm a}$ & & \protect\cite{Faure2008}\\ 
COSJ100013+023424 &I& 150.055775328 & 2.573347012 & 1.45 & 0.890$^{\rm b}$ && This work\\ 
COSJ100018+022138 &J& 150.076934522 & 2.360757128 & 0.38 & 1.53$^{\rm c}$ & 3.42$^{\rm c}$ & \protect\cite{vanderWel2013} \\ 
COSJ100024+015334 &K& 150.100467215 & 1.893029318 & 0.77 & 2.02$^{\rm b}$ & 5.10$^{\rm d}$ & \protect\cite{Mercier2024} \\ 
COSJ100024+021749 &L& 150.100088848 & 2.297132757 & 0.73 & 0.362$^{\rm a}$ & 2.63$^{\rm e}$ & \protect\cite{Jin2018, Jin24} \\ 
COSJ100025+015245 &M& 150.106707900 & 1.879287103 & 0.53 & 2.449$^{\rm b}$ & & This work\\ 
COSJ100028+021928 &N& 150.119067171 & 2.324503234 & 0.66 & 0.604$^{\rm a}$ & & This work\\ 
COSJ100047+015023 &O& 150.198584092 & 1.839784559 & 1.62 & 0.893$^{\rm a}$ & & \protect\cite{Faure2008} \\ 
COSJ100119+014849 &P& 150.330525811 & 1.813616188 & 1.14 & & &This work\\ 
COSJ100121+022740 &Q& 150.341363020 & 2.461297053 & 0.86 & 0.371$^{\rm a}$ & & This work\\ 
\hline
\end{tabular}
\caption{The 17 spectacular lenses from the COWLS sample revealed by our visual inspection, ordered by Right Ascension. The first column gives the ID from the larger catalogue published in \protect\cite{Shuntov2024}. The second column gives the ID letter matching \autoref{fig:JWSTvsHST}. The third and fourth column show the Right Ascension and Declination of the lenses respectively.  The fifth column gives the Einstein radius of the lens derived from the lens modelling performed in \protect\cite{Nightingale2025}. The sixth column gives the lens redshift where known, where $^{\rm a}$ denotes a spectroscopic redshift from DESI \protect\citep{Dey2019}, $^{\rm b}$ denotes a photometric redshift from the COSMOS-Web catalogue \citep{Shuntov2024} and $^{\rm c}$ denotes a spectroscopic redshift from \protect\cite{vanderWel2013}. The seventh column gives the source redshift where known, where $^{\rm d}$ denotes a spectroscopic redshift from \protect\cite{Shuntov2025} and $^{\rm e}$ denotes a spectroscopic redshift from \protect\cite{Jin24}. The last column gives references to systems that were previously known from other wavelength surveys.}
\label{tab:lens_table}
\end{table*}

\section{Discussion and conclusions} \label{sec:discussion}

This work presents $17$ spectacular lenses within the COSMOS-Web field selected based on the visual inspection of $42\,660$ objects observed with NIRCam.  Our first result is therefore that \textit{JWST}NIRCam imaging can not only find strong lenses, but reveal spectacular ones that lensing experts unanimously agree are lenses without spectroscopic follow-up, even within a small $0.54$\,deg$^2$ patch of the Universe.

\begin{figure*}
    \centering
    \includegraphics[width=\textwidth,trim={0 13.0cm 0 0},clip]{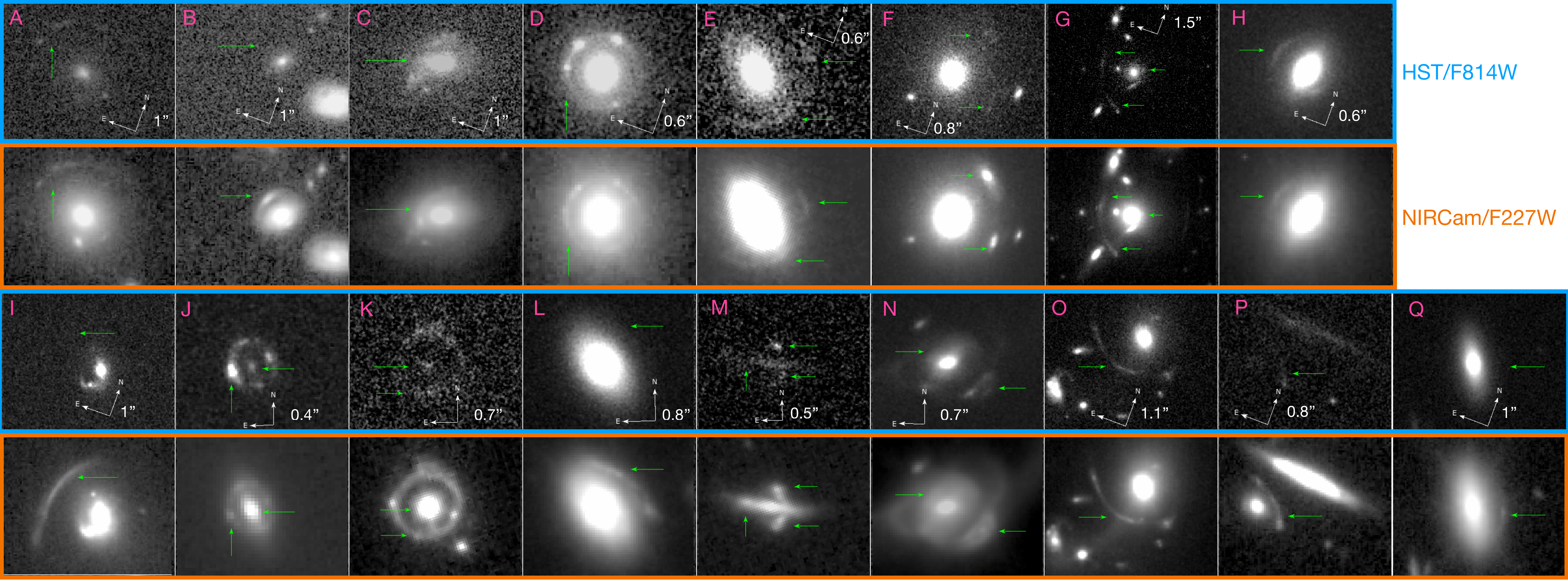}
    \caption{This figure shows the comparison of features from the \textit{HST} ACS/F814W filter and the \textit{JWST} NIRCAM/F277W filter for the 17 lenses presented in this work. We emphasise the power of \textit{JWST} images to reveal clear evidence of lensing, showing how this observatory has the capacity to discover new, spectacular lenses. So as not to clutter the image we used a letter designation for each target, reported in \autoref{tab:lens_table}. 
    }
    \label{fig:JWSTvsHST}
\end{figure*}


We visually compare the NIRCam images of our lenses with previous \textit{HST}/F814W coverage in \autoref{fig:JWSTvsHST}. As detailed in \autoref{tab:lens_table}, some of the lenses detected in our visual inspection have previously been identified as potential lens candidates using \textit{HST} data, and other instruments. One lens, labelled J, was previously confirmed with \textit{HST} \citep{vanderWel2013} and shows clear evidence for lensing in both \textit{HST} and JWST data. The lens L was confirmed using submillimetre and far-infrared deblending techniques \citep{Jin2018} and is not detected in \textit{HST} (or NIRCam/F115W imaging, which is $\sim1$ mag deeper). The lens G was identified by \citet{More2012} using ground-based Canada–France–Hawaii Telescope Legacy Survey imaging and is clearly visible in both \textit{HST} and \textit{JWST}. 

Lenses A, D, E, H, and O were candidates in previous search based on \textit{HST} images \citep{Faure2008, Pourrahmani2018}, with classifications ranging from convincing lenses (D and O) to more subjective ones based on \textit{HST} data alone. Of the 9 new COWLS candidates (including the COSMOS-Web ring), the source emission in six of them (C, F, M, N, K, and P) is detected in \textit{HST} data, but it was insufficient to classify them as lens candidates in prior lens searches. Our sample therefore includes four \textit{HST}-dark lensed galaxies \citep{Perez-Gonzalez2023}, labelled B, I, L, and Q. Nevertheless, we stress that the resolution of NIRCam imaging is a critical factor which led to the confirmation of lensing features in our inspection. The lens COSJ095920+015851 (labelled C in \autoref{fig:JWSTvsHST}) exemplifies this; in the \textit{HST} image the lensed arc is apparent but poorly resolved, whilst in the \textit{JWST} image the arc is well resolved and unambiguously the product of strong lensing.

Furthermore, the multi-wavelength nature of the COSMOS-Web survey means that lenses can be unambiguously selected based on the distinct colour differences between the lower-redshift lenses and higher-redshift sources. We emphasise that the combination of depth, resolution and near-infrared coverage provided by NIRCam allows for clear identification of lensing features that were in the past ambiguous, thanks to well-resolved spatial separations between lens galaxies and their respective lensed source light, and the colour separation enabled by the multi-wavelength coverage.

\begin{figure*}
\centering
\includegraphics[width=0.19\textwidth]{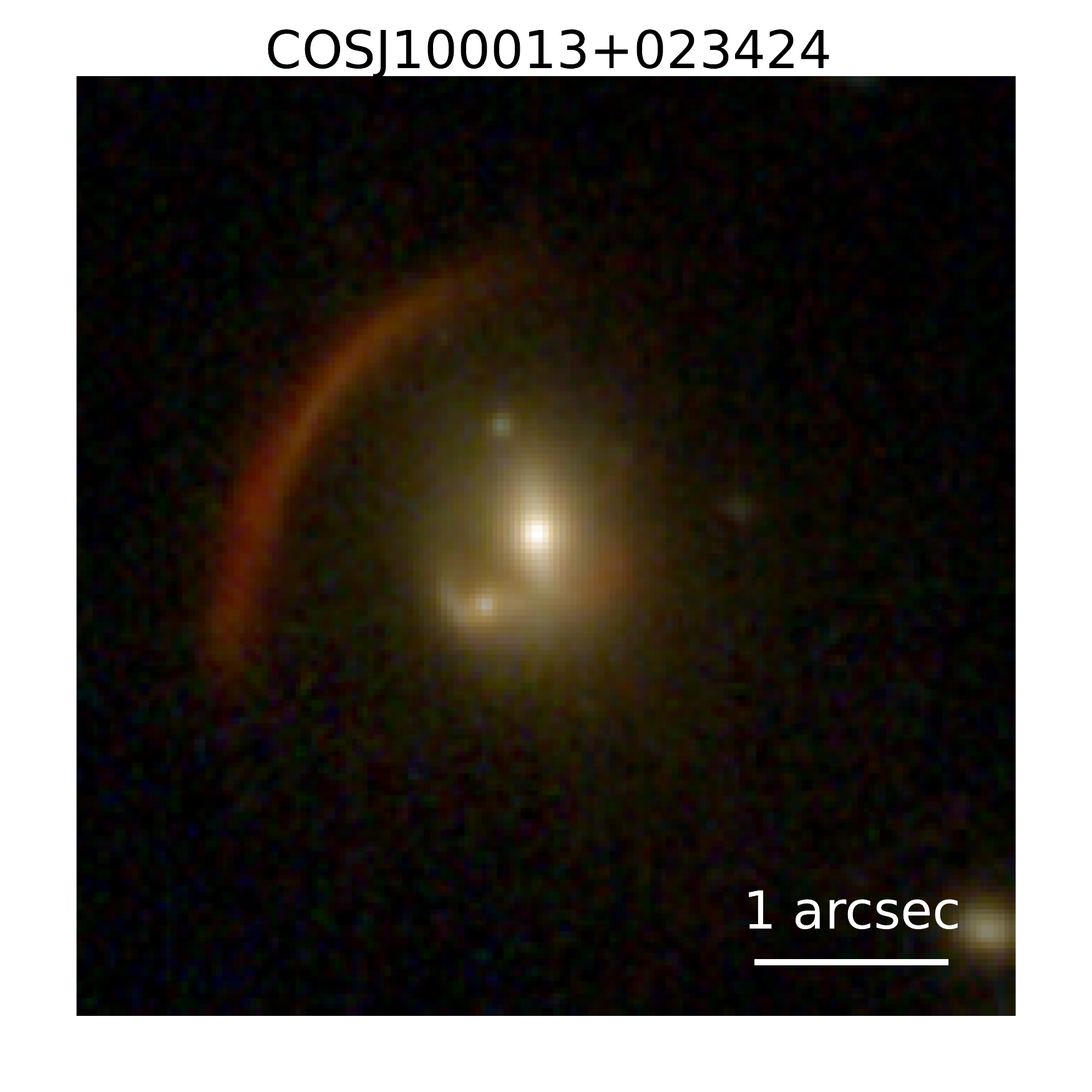}
\includegraphics[width=0.19\textwidth]{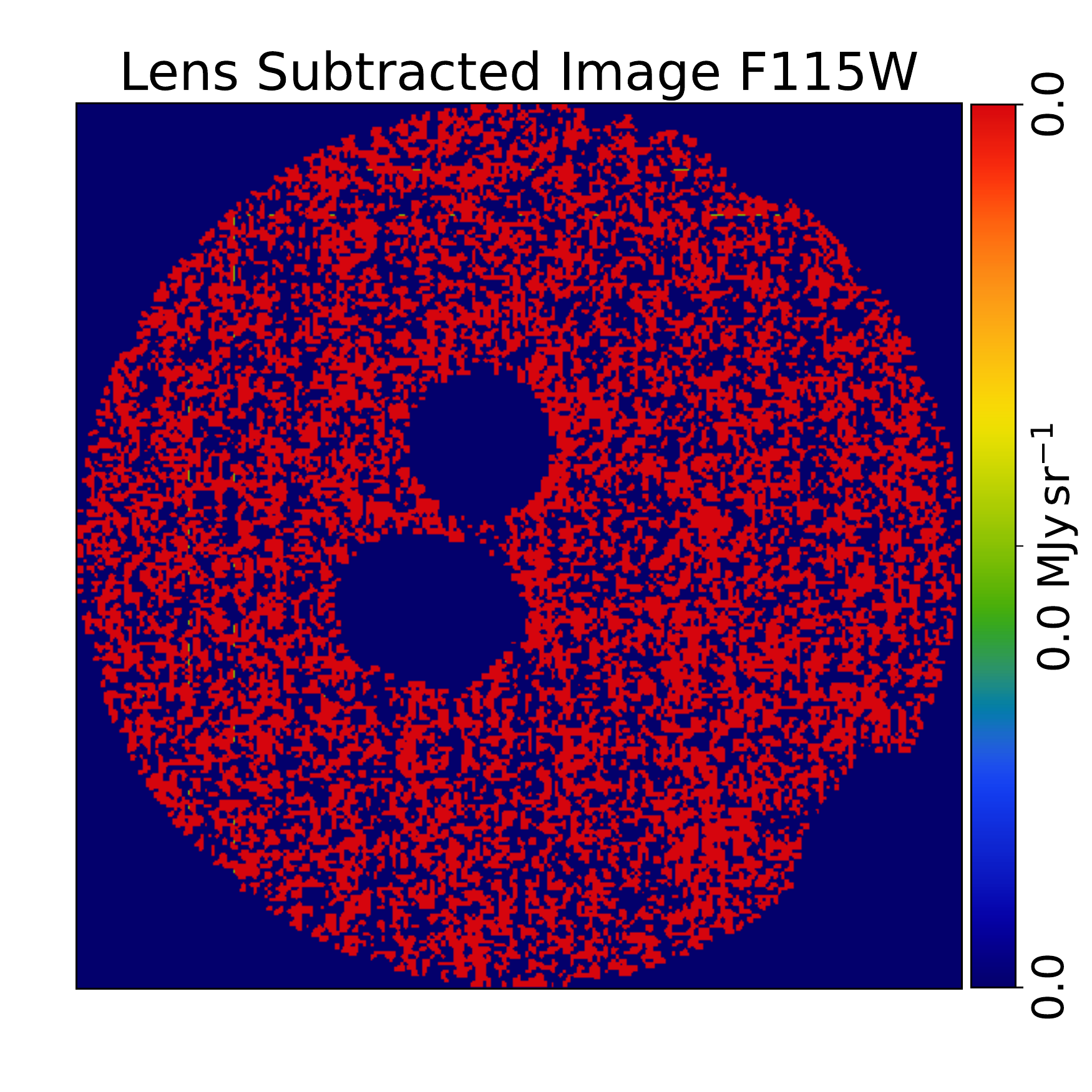}
\includegraphics[width=0.19\textwidth]{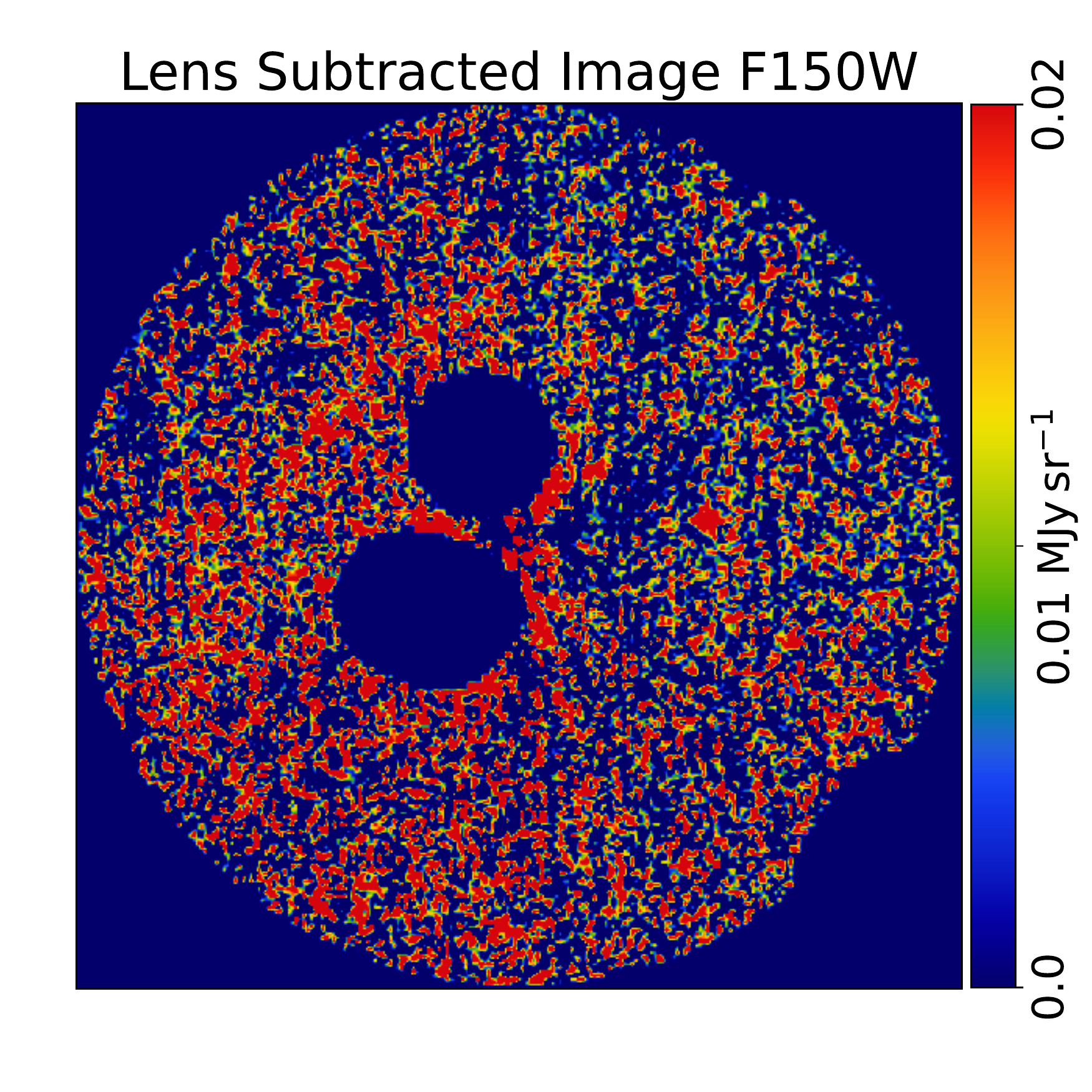}
\includegraphics[width=0.19\textwidth]{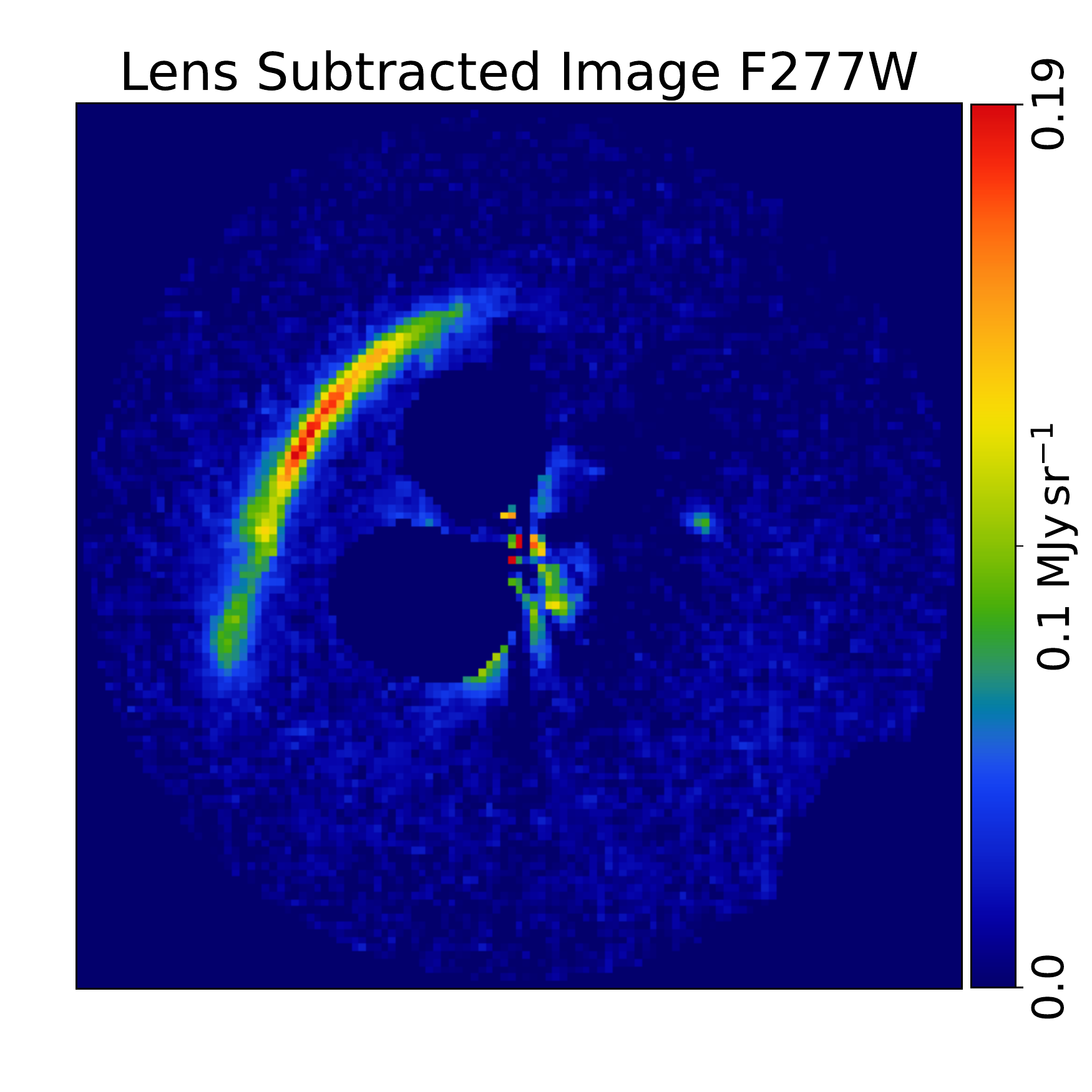}
\includegraphics[width=0.19\textwidth]{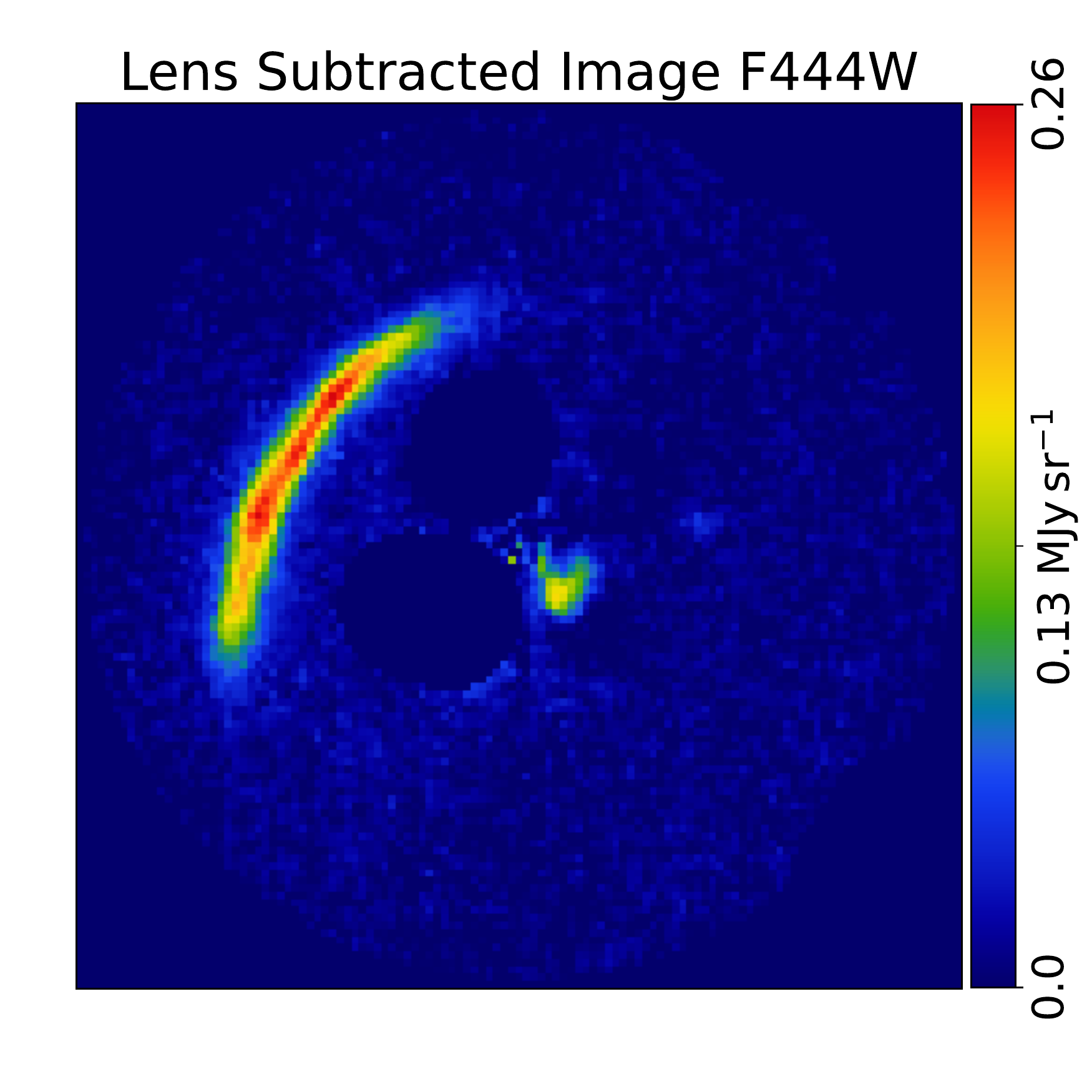}
\includegraphics[width=0.19\textwidth]{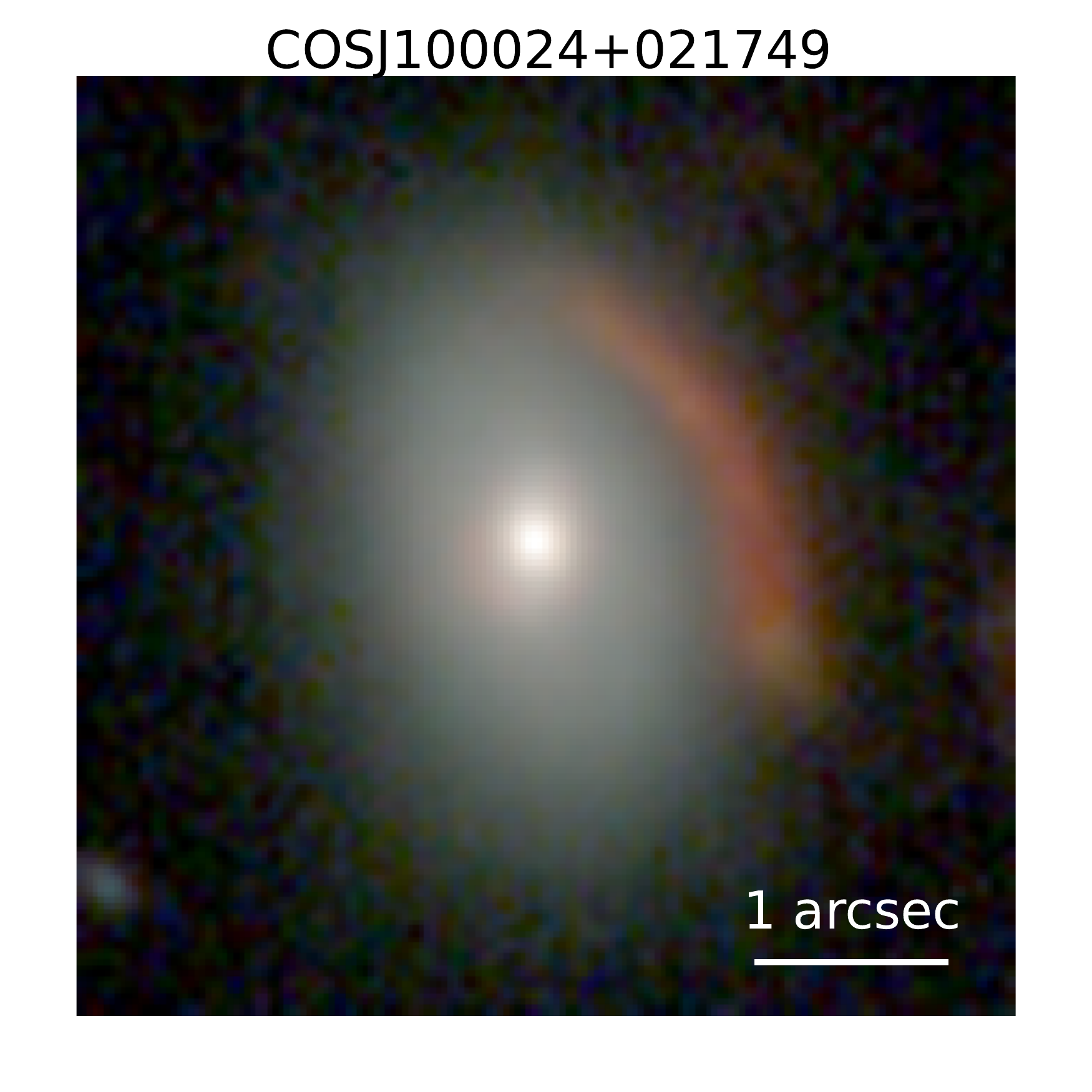}
\includegraphics[width=0.19\textwidth]{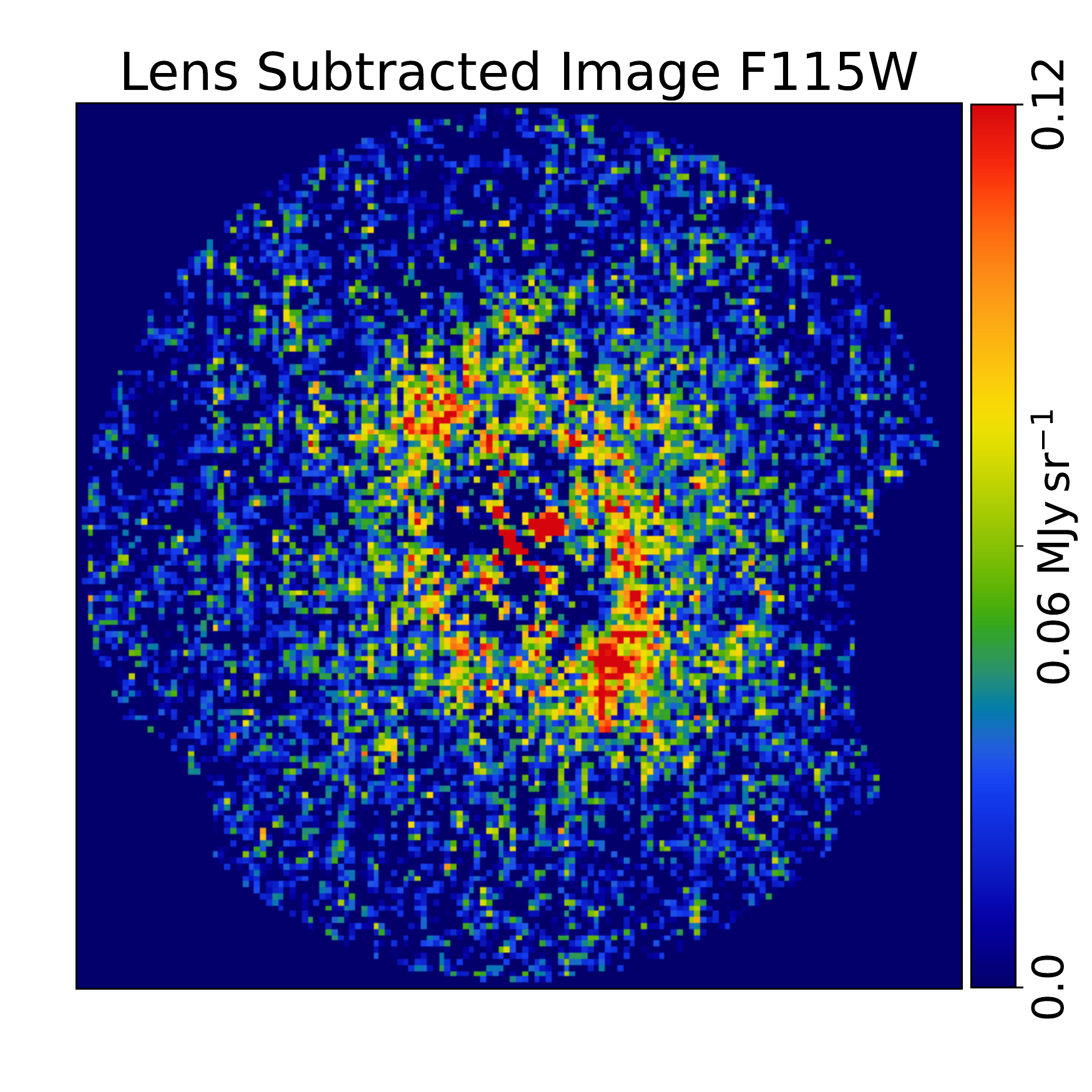}
\includegraphics[width=0.19\textwidth]{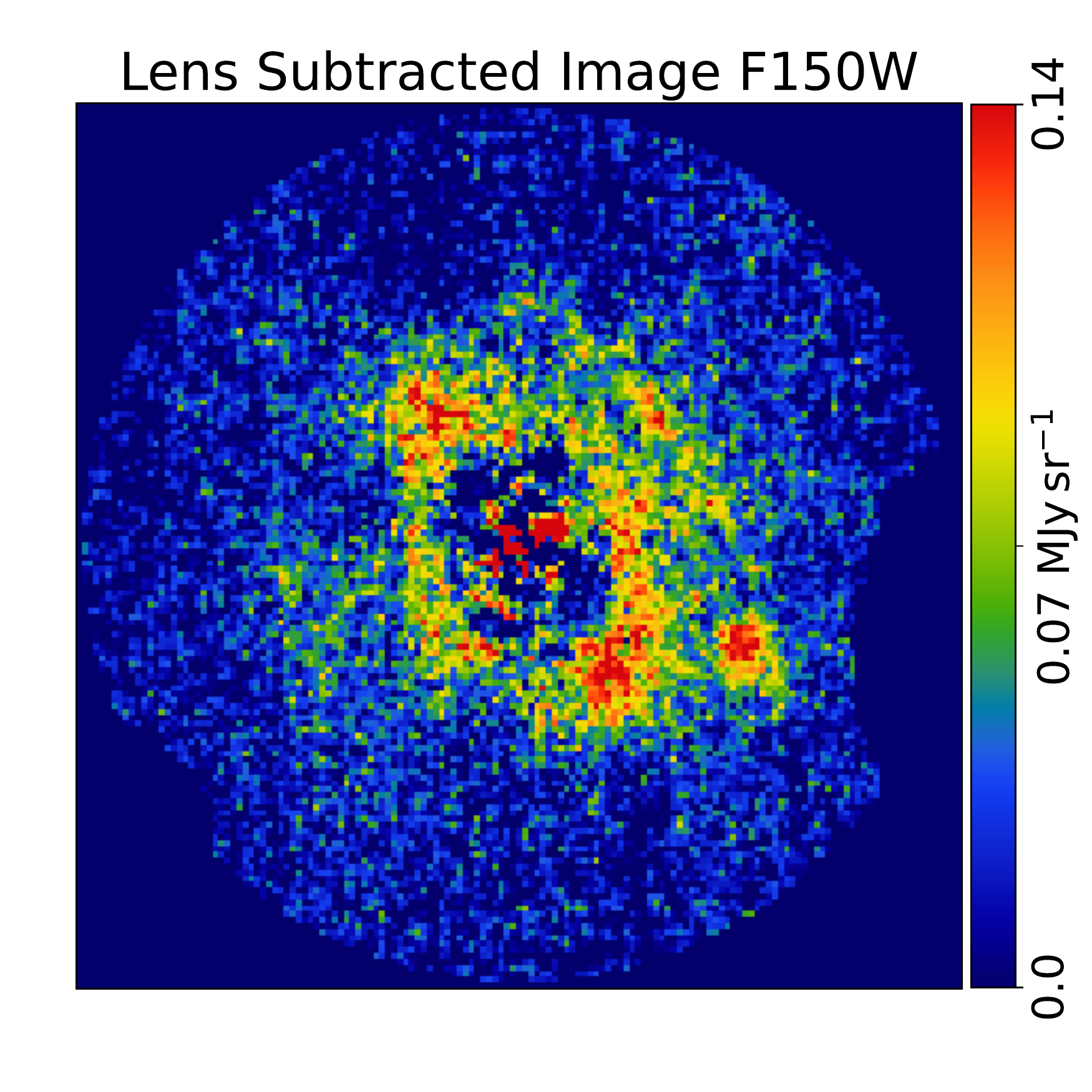}
\includegraphics[width=0.19\textwidth]{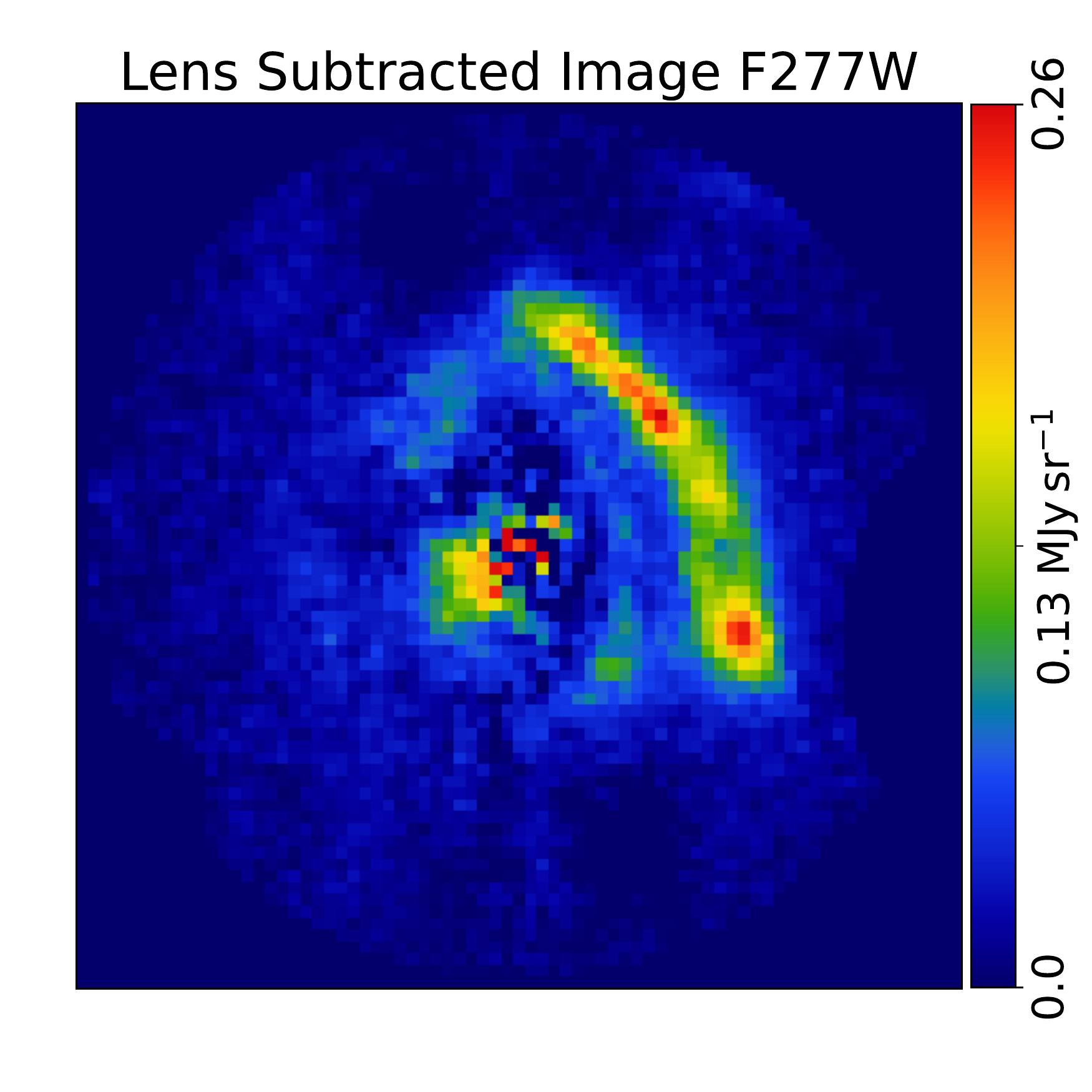}
\includegraphics[width=0.19\textwidth]{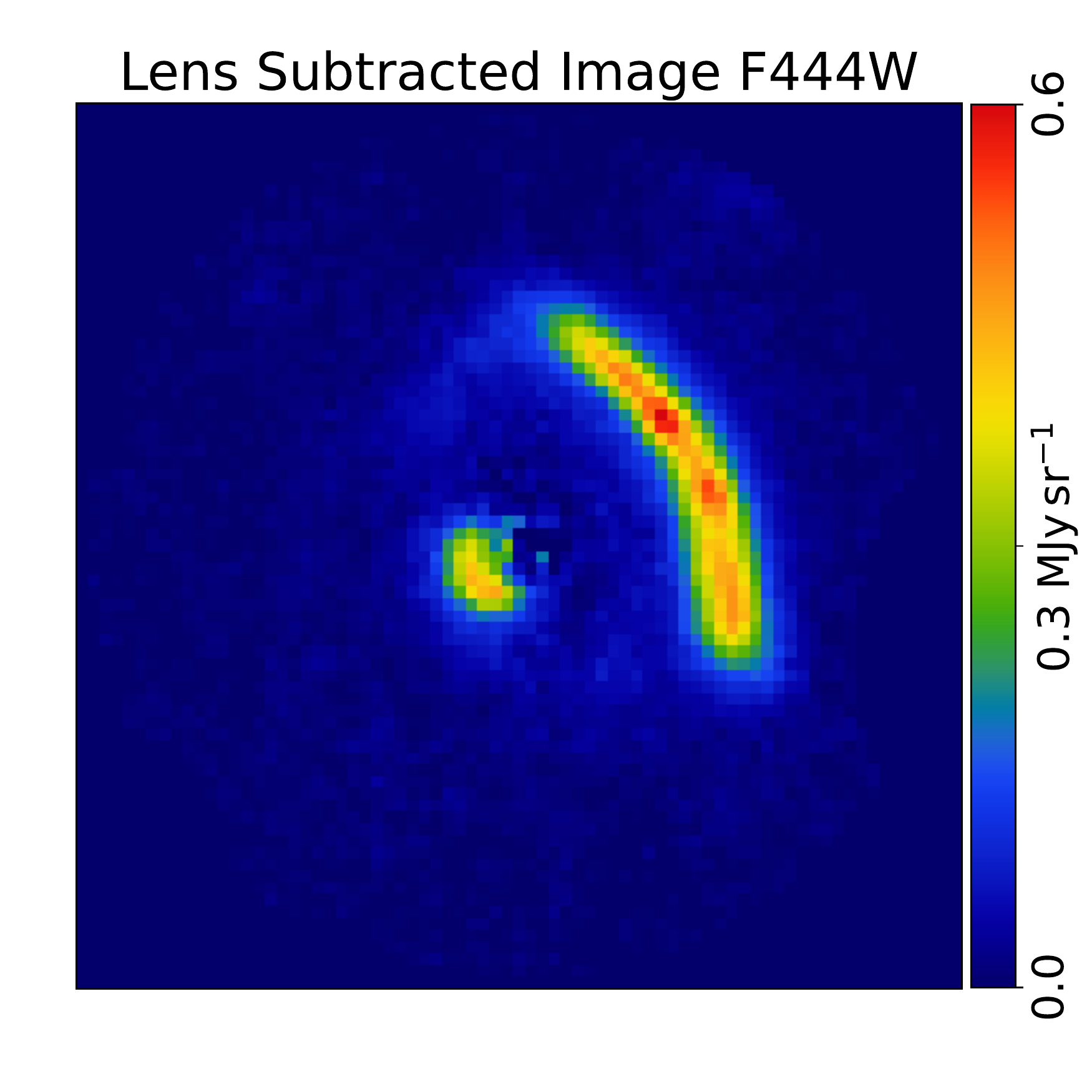}
\includegraphics[width=0.19\textwidth]{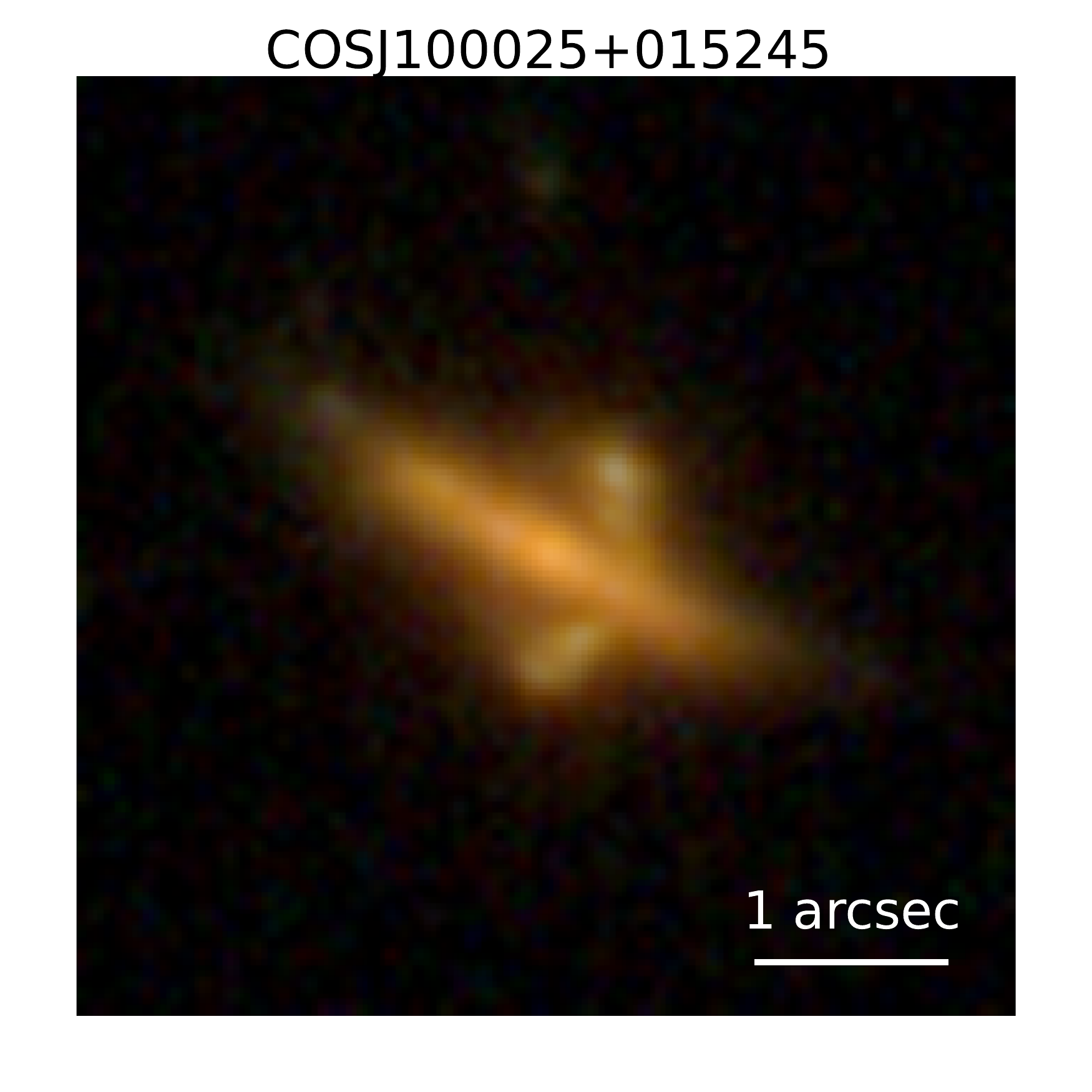}
\includegraphics[width=0.19\textwidth]{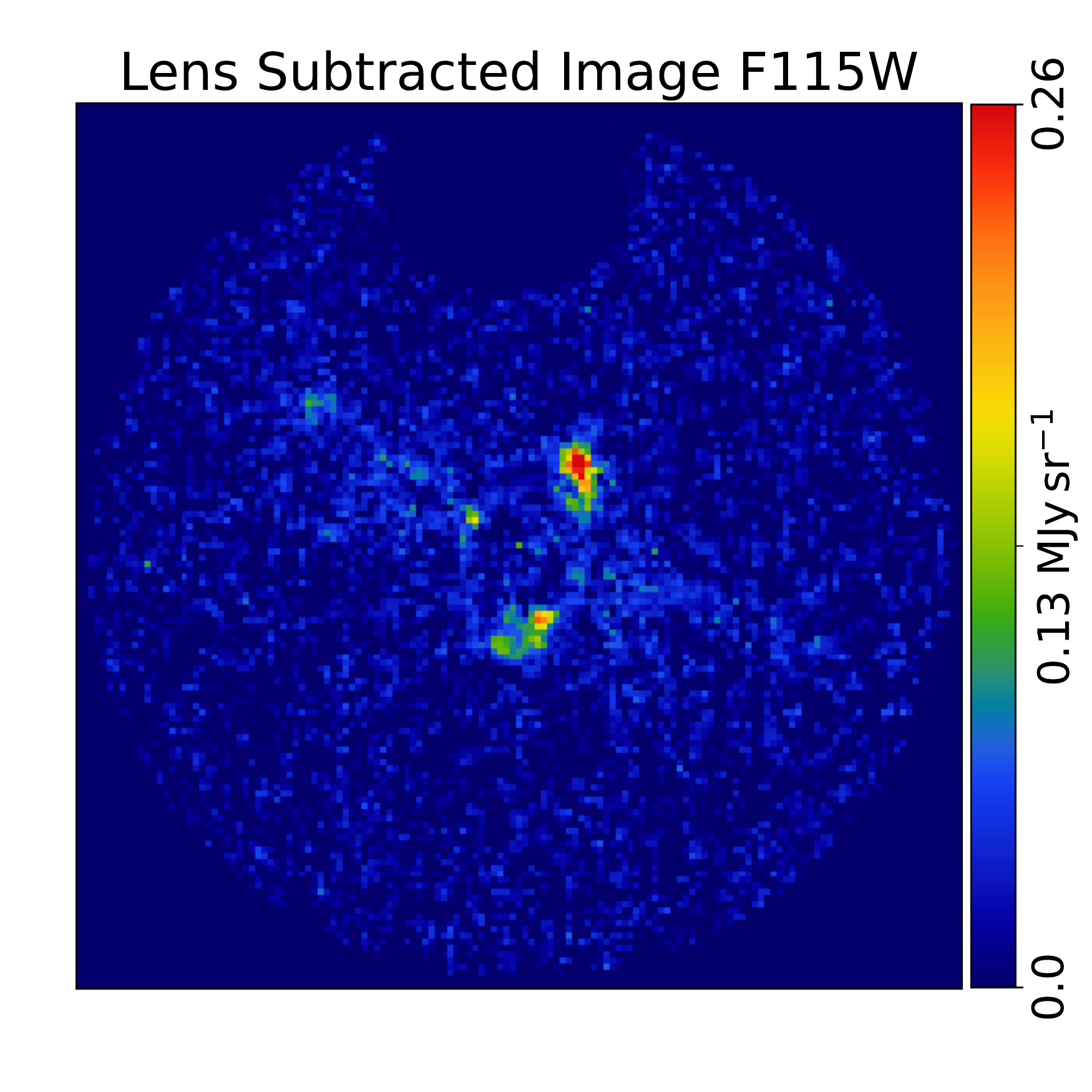}
\includegraphics[width=0.19\textwidth]{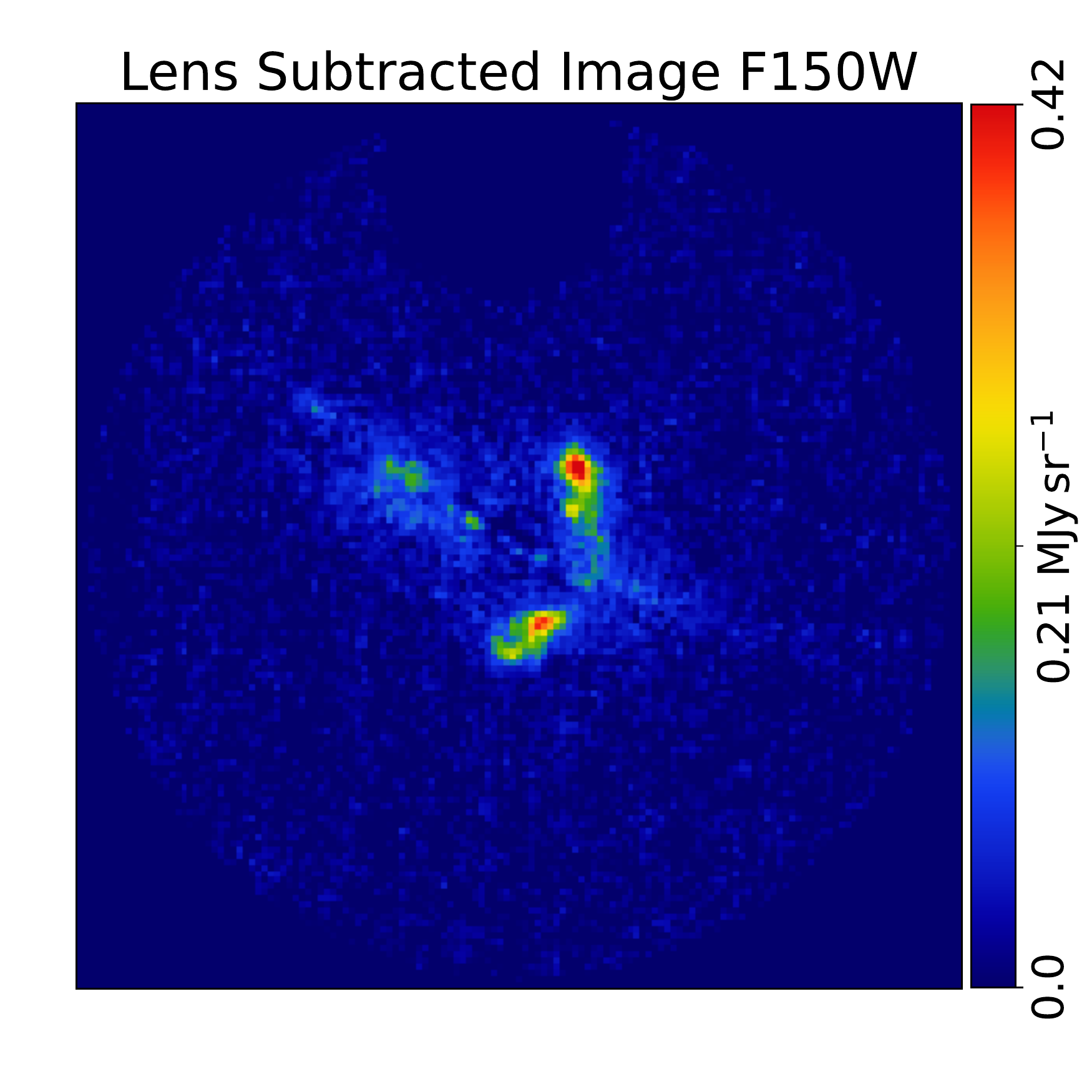}
\includegraphics[width=0.19\textwidth]{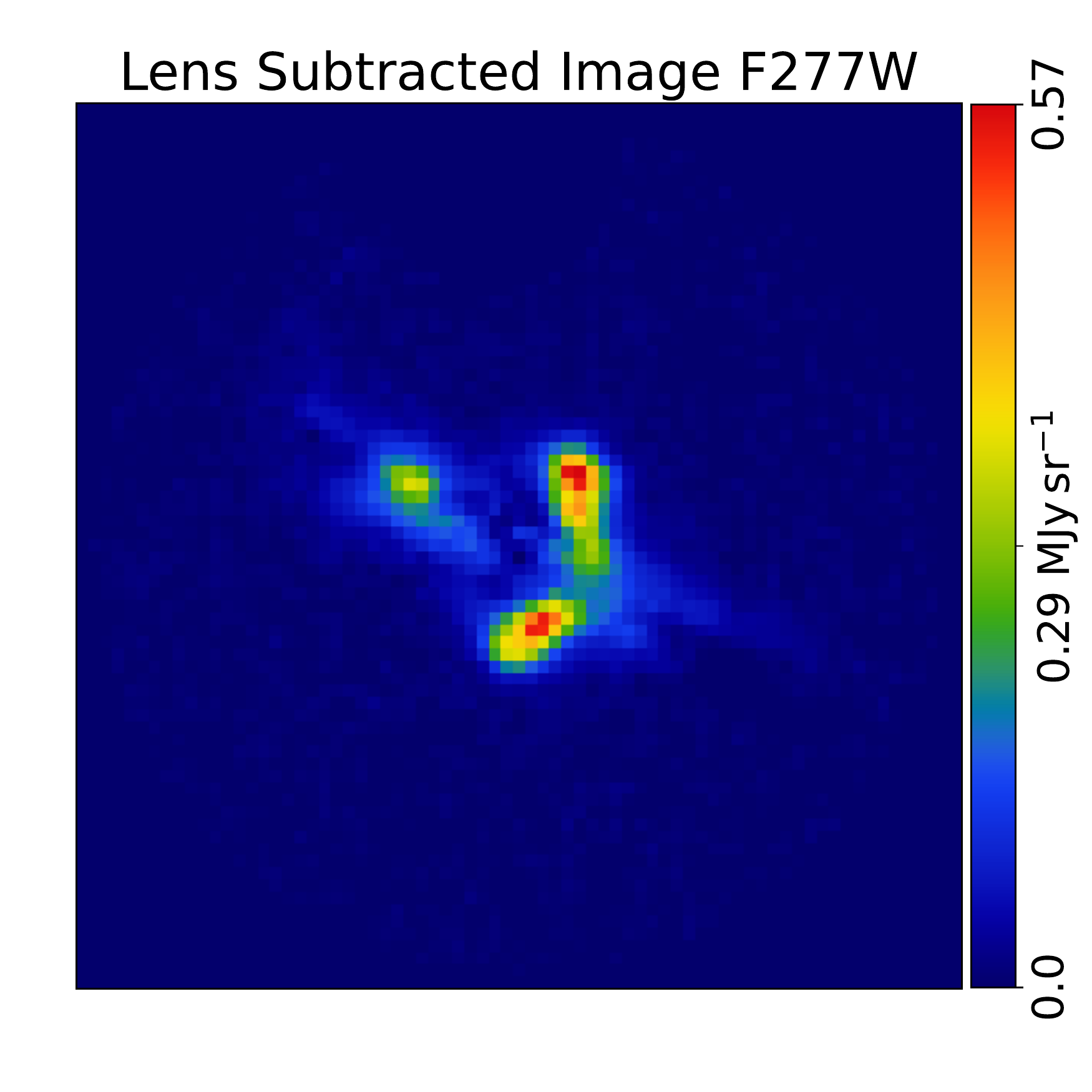}
\includegraphics[width=0.19\textwidth]{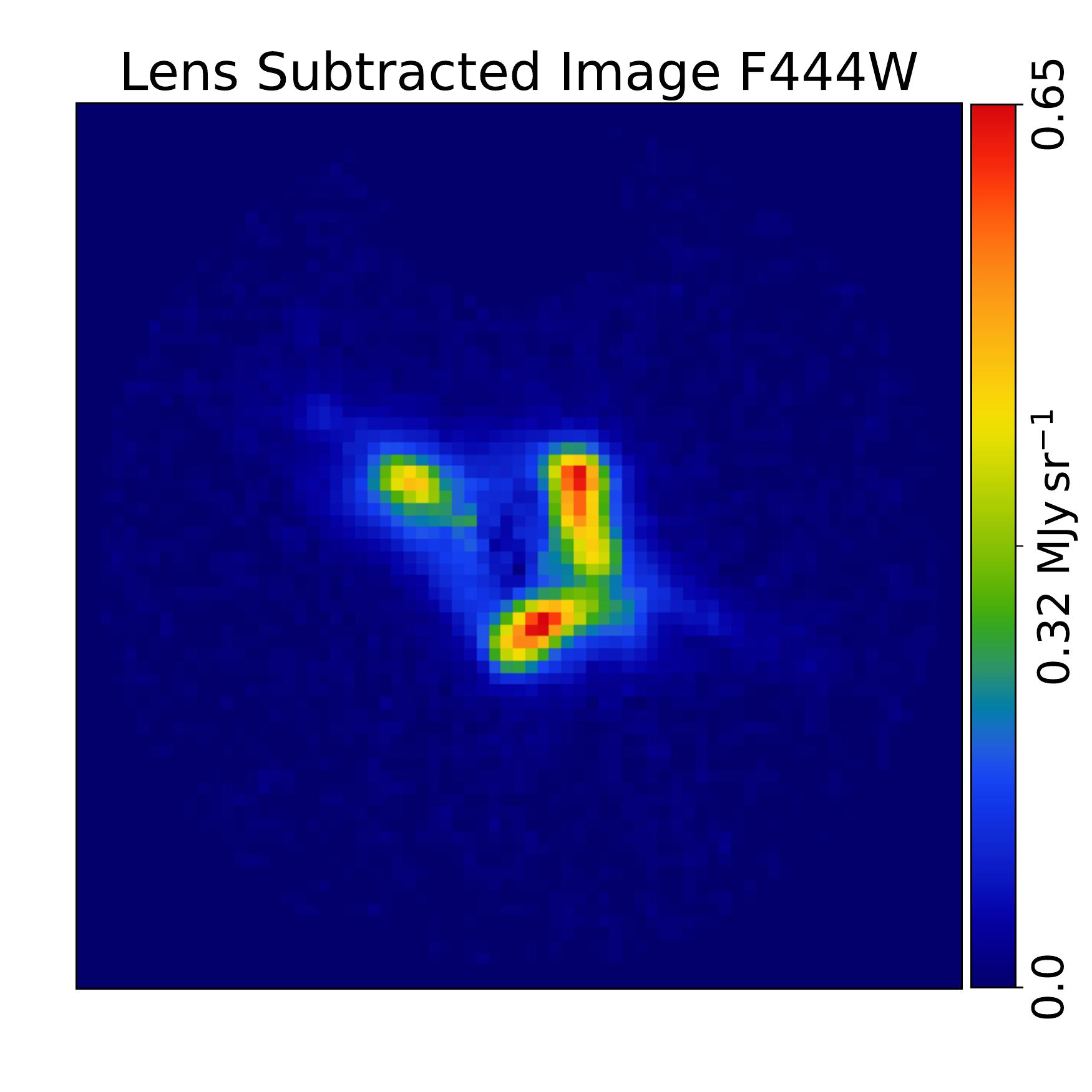}
\caption{
Three illustrative examples demonstrating how the high S/N multi-wavelength imaging of \textit{JWST} provides new insights for lens modelling. The left column shows RGB images constructed from all four wavebands, the next four columns display lens-subtracted images for individual wavebands (F115W, F150W, F277W, F444W), using \texttt{PyAutoLens} lens models from COWLS Paper I.  The three example lenses from top to bottom illustrate:  (i) \textit{COSJ100013+023424}: how HST-dark sources are also invisible in shorter wavelength F115W and F150W imaging, leaving only the lens light visible and making lens and source deblending easier. (ii) \textit{COSJ100024+021749}: how the source morphology may change significant across different wavelengths, where in this lens at F115W / F150W wavelengths a quadruply imaged ring is observed compared to cusp like giant arc at F277W / F444W wavelengths. Combining multi-wavelength datasets like this will provide significantly more constraints on the lens mass model.  (iii) \textit{COSJ100025+015245}: how shorter-wavelength emission from the source can be absorbed by dust in the lens galaxy.  
}

\label{figure:LensModel}
\end{figure*}

Owing to their multi-wavelength discovery and selection, these spectacular lenses enable unique scientific studies and provide a wealth of additional information for lens modeling. \autoref{figure:LensModel} illustrates this for three example lenses, showing their RGB images alongside individual lens-subtracted images in each waveband (F115W, F150W, F277W, F444W). \textit{HST}-dark sources result in the source being undetected at bluer wavelengths, leaving only the lens visible. This allows for a much cleaner deblending of lens and source light than is possible in \textit{HST} lens datasets. For many spectacular lenses, the source morphology changes dramatically across wavelengths, altering the observed image-plane structure and providing significantly more constraints on the lens mass model. Additionally, dust absorption in the source can be directly observed when comparing images from redder to bluer wavelengths, offering insights not previously accessible.

This finding initiates a new era of strong lensing observations. We have shown how NIRCam coverage, even shallow,  can reveal spectacular lenses at an unprecedented rate. We can extrapolate our findings of 17 spectacular lenses in the $0.54$ deg$^2$ COSMOS-Web field to other \textit{JWST} surveys by matching this area to the size of the NIRCam field-of-view. We find that every 10 to 12 NIRCam pointings will reveal a spectacular strong lens, according to whether or not the chip gap between Module A and B is covered. This means that survey such as BEACON should yield six spectacular lenses in its $0.2$ deg$^2$ area; CEERS and JADES will contain three each in their $0.1$ deg$^2$ areas.

Going beyond the spectacular lenses exhibited here, the overall prediction of about 100 lenses initially estimated in COSMOS-Web field (\citealt{Casey2023}; see also \citealt{Holloway:2023axl, Ferrami:2024obm}), are well matched by the full COWLS sample (COWLS Paper I, \citealt{Nightingale2025}). The estimation of \cite{Casey2023} was a simple calculation based on previous visual inspection efforts of \textit{HST} observations in the same field. In COWLS Paper III, \cite{Hogg2025b}, we present a simulation-based estimate of lens abundance and properties in the COSMOS-Web field, and compare the result with the observed COWLS lenses. This paves the way for the characterisation of the selection function of the sample. 

In conclusion, this work presents the 17 most spectacular strong gravitational lenses identified through visual inspection of COSMOS-Web data, alongside the full COWLS catalogue of over 100 high-confidence candidates. These lenses feature some of the highest-redshift sources ($z > 6$) and lenses ($z > 2$). They reside within a compact $0.54$ deg$^2$ region enabling joint strong and weak lensing analysis, and include systems with small lens-source separations that may enable supermassive black hole mass measurements. The use of multi-wavelength NIRCam data was key to identifying these spectacular and unique systems, with the promise of more such lenses to be discovered in all future NIRCam observations.

\section*{Acknowledgements}
We thank Kaija Virolainen for her contribution to the the image rendering.
JWN is supported by an STFC/UKRI Ernest Rutherford Fellowship, Project Reference: ST/X003086/1. QH acknowledges support from the European Research Council (ERC) through Advanced Investigator grant DMIDAS (GA 786910). This work was made possible thanks to the CANDIDE cluster at the Institut d’Astrophysique de Paris, which was funded through grants from the PNCG, CNES, DIM-ACAV, and the Cosmic Dawn Center; CANDIDE is maintained by S. Rouberol. The French contingent of the COSMOS team is partly supported by the Centre National d'\'Etudes Spatiales (CNES). OI acknowledges the funding of the French Agence Nationale de la Recherche for the project iMAGE (grant ANR-22-CE31-0007). SJ acknowledges the Villum Fonden research grants 37440 and 13160 DS acknowledge the Jet Propulsion Laboratory, California Institute of Technology, under a contract with the National Aeronautics and Space Administration (80NM0018D0004).

\section*{Data Availability}
The data underlying this article are available in the COWLS repository \href{https://github.com/Jammy2211/COWLS_COSMOS_Web_Lens_Survey}{\faGithub}.

\bibliographystyle{mnras}
\bibliography{paper, james} 

\noindent\rule{\columnwidth}{0.4pt}
$^{1}$\Liege\\
$^{2}$\DurhamCEA\\
$^{3}$\DurhamICC\\
$^{4}$\Newcastle\\
$^{5}$\LUPM\\
$^{6}$\Aalto\\
$^{7}$\Helsinki\\
$^{8}$\Northeastern\\
$^{9}$\UTAustin\\
$^{10}$\PMO\\
$^{11}$\LAM\\
$^{12}$\JPL\\
$^{13}$\DAWN\\
$^{14}$\NBI\\
$^{15}$\IAP\\
$^{16}$\LAM\\
$^{17}$\UCSB\\
$^{18}$\Rochester\\
$^{19}$\STScI\\
$^{20}$\UCSC\\
$^{21}$\NAOJ\\
$^{22}$\Hawaii\\
$^{23}$\Caltech\\
$^{24}$\DTU\\

\bsp	
\label{lastpage}
\end{document}